\documentclass[traditabstrac]{aa} 

\newcommand{\micron}{$\mu$m}

\usepackage{graphicx}
\usepackage{txfonts}
\usepackage{mathtools}  
\usepackage{colortbl}

\usepackage{graphicx}
\usepackage{txfonts}
\usepackage{natbib}
\begin{document}

   \title{Measuring elemental abundance ratios in protoplanetary disks at millimeter wavelengths}


   \author{D. Fedele
          \inst{\ref{inst:inaf_oaa}}
          \and
          C. Favre\inst{\ref{inst:ipag}}
          }

   \institute{
   Istituto Nazionale di Astrofisica, Osservatorio Astrofisico di Arcetri, L.go E. Fermi 5, 50126, Firenze (Italy)\label{inst:inaf_oaa}\\
              \email{davide.fedele@inaf.it}
         \and
 Univ. Grenoble Alpes, CNRS, IPAG, F-38000 Grenoble, France\label{inst:ipag}\\
                \email{cecile.favre@univ-grenoble-alpes.fr}
             }
   \date{Received ...; accepted ...}

 \abstract{
Over millions years of evolution, gas dust and ice in protoplanetary disks can be chemically reprocessed. There is evidence that the gas-phase carbon 
and oxygen abundances are subsolar in disks belonging to nearby star forming regions. These findings have a major impact on the composition of the primary 
atmosphere of giant planets (but it may also be valid for super-Earths and sub-Neptunes) as they accrete their gaseous envelopes from the surrounding material
in the disk.  In this study, we performed a thermochemical modeling analysis with the aim of testing how reliable and robust are the estimates 
of elemental abundance ratios based on (sub)millimeter observations of molecular lines. We created a grid of disk 
models for the following different elemental abundance ratios: C/O, N/O, and S/O, and we computed the line flux of a set of carbon-nitrogen and sulphur-bearing species, namely CN, HCN, NO, C$_{2}$H, c--C$_{3}$H$_{2}$, H$_{2}$CO, HC$_{3}$N, CH$_{3}$CN, 
CS, SO, H$_{2}$S, and H$_{2}$CS, which have been detected with  present (sub)millimeter facilities such as ALMA and NOEMA.
We find that the line fluxes, once normalized to the flux of the $^{13}$CO $J=2-1$ line, are sensitive to the elemental abundance
ratios.  On the other hand, the stellar and disk physical parameters have only a minor effect on the line flux ratios. Our results 
demonstrate that a simultaneous analysis of multiple molecular transitions is a valid approach to constrain the elemental abundance
ratio in protoplanetary disks.
}

   \maketitle
%

\section{Introduction}
Planets inherit their chemical composition from the protoplanetary disk in which they form.
Because of thermal and chemical reprocessing, the relative abundance of different species 
within a disk can differ from the values of the natal molecular cloud. 
There is, in fact, evidence of nonsolar abundance of gas-phase carbon in some protoplanetary disks:
TW Hya (e.g., \citealt{Bergin13, Favre13, Kama16}), GM Aur \citep{Mcclure16}, DM Tau \citep{Mcclure16} 
and HD 100546 \citep{Kama16}.  Compared to the solar abundance of C/H = 2.69 $\times$ 10$^{-4}$ \citep{Asplund09}, 
these systems appear to have subsolar carbon abundance. In particular, TW Hya and GM Aur show 
a substantial deficit by nearly 2 orders of magnitude. These results are based on the detection of deuterated hydrogen (HD)
emission with the Herschel Space Observatory (see e.g., \citealt{Bergin13, Kama20}), which allows us to derive robust constraints on overall gas mass of disks. 
Further indications of the under abundance of gas-phase carbon in disks come from recent ALMA observations of multiple CO
isotopologues, which reveal a surprisingly low CO abundance in the disk population of the Lupus star forming region 
\citep[e.g.,][]{Ansdell16, Miotello17, Zhang20a}. 

\smallskip
\noindent
The elemental abundance ratios within a disk is relevant for the composition of the primary gaseous atmosphere of giant planets. 
Several authors have investigated how the distribution of ices and volatiles in disks affect the final composition of planets. Growing 
attention is given to the gas-phase elemental abundance ratio of carbon-to-oxygen (hereafter C/O) in disks (e.g., \citealt{Oberg11, Piso15,
Mordasini16, Espinoza17, Madhusudhan17, Cridland19}). In this regard, there is evidence of disk-to-disk scatter of the gas-phase
C/O abundance ratios (e.g., \citealt{Semenov18, Cleeves18}).

\smallskip
\noindent
This paper presents a modeling analysis that aims to test the possibility of measuring the global values of the 
elemental abundance ratios C/O, N/O, and S/O in disks by means of (sub)millimeter observations of molecular transitions.
The model details are given in Section~\ref{sec:model}, and the results of the grid are presented and discussed in Sections~\ref{sec:results} and 
\ref{sec:discussion}, respectively. Conclusions are given in Section~\ref{sec:conclusions}.    


\begin{figure*}[!th]
        \centering
        \includegraphics[width=18cm]{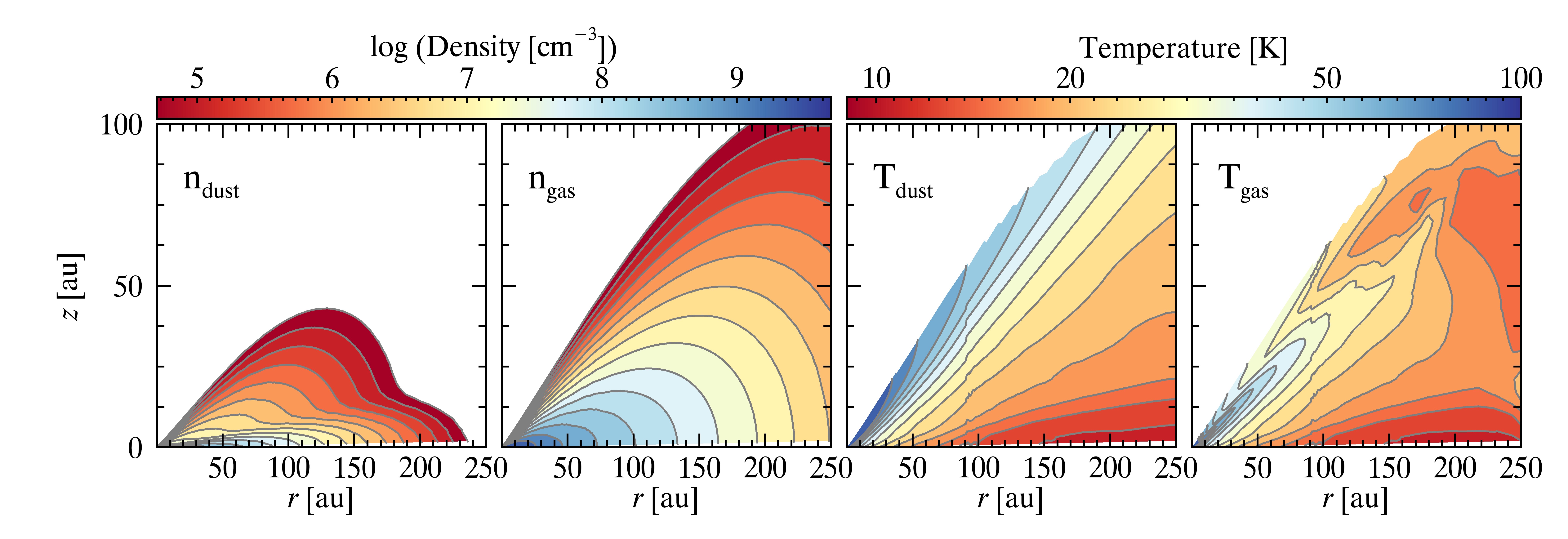}
        \caption{Physical structure of the DALI reference model (boldface values in Table~\ref{tab:setup}) showing the density and temperature structure of dust and gas.}\label{fig:dali1}
\end{figure*}

\begin{table}
        \begin{tabular}{lcl}
                \hline\hline
                Parameter & Value & Description\\
                \hline
                $M_{*}$ [M$_{\odot}$]   & 0.9        & Stellar mass                  \\
                $T_{eff}$ [K]           & 3250       & Stellar temperature \\
                $L_{*}$ [L$_{\odot}$]   & {\bf 1.5}; 5; 10 & Stellar luminosity            \\
                $L_{X}$ [erg\,s$^{-1}$] & {\bf 10$^{29}$}; $10^{30}$; $10^{31}$  & Stellar X-rays luminosity     \\
                $R_{in}$ [au]           & 0.15       & Inner disk radius             \\
                $R_{out}$ [au]          & 300.0      & Outer disk radius             \\
                $\gamma$                & 0.2        & $\Sigma(r)$ power-law exponent \\
                $\Sigma_c$              & 1.0        & $\Sigma(r)$ at $R=R_c$        \\
                $R_c$ [au]              & 120.0                  & Tapering disk radius          \\
                $h_c$                   & 0.08; {\bf 0.13}; 0.15 & Disk scale height at $R_c$    \\
                $\psi$                  & 0.0; {\bf 0.1}; 0.2  & Disk flaring exponent         \\
                $\Delta_{gd}$           & 5;   {\bf 10};  50   & Gas-to-dust mass ratio        \\
                $p$                                             & 3.5                      & Dust power-law exponent  \\
                $a_{min}$ [\micron]   & 0.001; {\bf 0.005}; 0.01  & Minimum grain size\\ 
                $\chi$                & {\bf 0.2}, 0.4, 0.6                     & Dust settling      \\
                $f_{large}$       & 0.55, {\bf 0.85}, 0.95            & Large grains mass fraction       \\
                \hline\hline
        \end{tabular}
        \caption{Input physical parameters. The values in boldface refer to the reference setup used for the model grid.}\label{tab:setup}
\end{table}

\section{DALI thermochemical disk models}\label{sec:model}
This work is based on simulations performed with the DALI thermo-chemical code of disks \citep{Bruderer12}. 
DALI takes the stellar spectrum and the disk density structure as input. The code
computes the dust temperature and radiation field strength by solving the dust continuum 
radiative transfer. In this work, the time-dependent thermochemistry is 
evaluated taking a typical age of 10$^6$\,yr. 
The dust-continuum and line-emission maps are finally estimated through ray tracing. 
The collisional rates are taken from the LAMDA database \citep{Schoier05}.
The chemical network used in this work is based on the UMIST database \citep{Woodall07} and it 
is made of 167 species and 2138 reactions. The calculation starts with atomic abundances 
(with all molecular abundances set to zero) including nine elements (H, He, C, N, O, S, Mg, Si, Fe). 
The list of reactions includes (besides gas phase reactions): H$_2$ 
formation on grains, freeze-out of molecules on grains, hydrogenation of ices, photodesorption, 
photodissociation, X-rays, cosmic-ray induced reactions, and PAH exchange charge reactions.

\subsection{Disk physical structure}
The gas surface density adopted here is described by a power-law radial profile with an exponential cut-off:

\begin{equation}
\Sigma_{\rm gas}(R) = \Sigma_{\rm c}  \ \Bigg(\frac{R}{R_{\rm c}}\Bigg)^{-\gamma} \ \exp\Bigg[ - \Bigg( \frac{R}{R_{\rm c}} \Bigg)^{2 - \gamma} \Bigg],
\end{equation}

\noindent
where $R_c$ is the cut-off radius and  $\Sigma_c$ the gas surface density at $R = R_c$.
The dust surface density is $\Sigma_{\rm gas}  / \Delta_{\rm gd}$, with $\Delta_{\rm gd}$ the gas-to-dust mass ratio.
In the vertical direction, the gas density is parameterized by a Gaussian distribution with scale height $h$ ($=H/R$) as follows:

\begin{equation}
h = h_{\rm c} \Bigg( \frac{R}{R_{\rm c}}\Bigg)^{\psi},
\end{equation}

\noindent
where $h_c$ is the gas scale height at $R = R_c$, and $\psi$ the degree of flaring.
Two populations of dust grains are included: small (size: 0.005 - 1\,\micron) and large (0.005 - 1000\,\micron), 
with a power-law size distribution and mass absorption cross sections as in \citet{Andrews11}. The small 
grains follow the same vertical distribution of the gas with scale height $h$, while the large grains have a 
reduced height $\chi h$ ($\chi < 1$) to account for the settling at the disk midplane. The dust surface density 
is $\Sigma_{\rm dust}  \ (1 - f_{\rm large})$  and $\Sigma_{\rm dust} \ f_{\rm large}$ for the small and large grains, 
respectively.  The adopted values are given in Table~\ref{tab:setup}.

\begin{table}[]
\centering
        \begin{tabular}{lllll}
                \hline\hline
                ID & N/H & O/H & C/O & N/O \\
                \#     & $\times 10^{-5}$ & $\times 10^{-4}$ & &  \\
                \hline
                1a& 0.21 & 28.8 & 0.047 & 7.4 $\times10^{-4}$  \\
                1b& 0.21 & 2.88 & 0.469 & 7.4 $\times10^{-3}$  \\
                1c& 0.21 & 1.35 & 1.000 & 1.6 $\times10^{-2}$  \\
                1d& 0.21 & 0.86 & 1.562 & 2.5 $\times10^{-2}$  \\               
                2a& 2.14 & 28.8 & 0.047 & 7.4 $\times10^{-3}$  \\
                2b& 2.14 & 2.88 & 0.469 & 7.4 $\times10^{-2}$  \\
                2c& 2.14 & 1.35 & 1.000 & 1.6 $\times10^{-2}$  \\
                2d& 2.14 & 0.86 & 1.562 & 2.5 $\times10^{-2}$  \\
                3a& 21.4 & 28.8 & 0.047 & 7.4 $\times10^{-2}$  \\
                3b& 21.4 & 2.88 & 0.469 & 7.4 $\times10^{-1}$  \\
                3c& 21.4 & 1.35 & 1.000 & 1.585                    \\
                3d& 21.4 & 0.86 & 1.562 & 2.477                    \\
                \hline\hline
                     & S/H & O/H &  S/O &  \\
                     & $\times 10^{-8}$ & $\times 10^{-4}$ & &  \\
                \hline
                2b-a& 0.19 & 2.88 &  6.6 $\times10^{-6}$ &  \\
                2b   & 1.91 & 2.88 &  6.6 $\times10^{-5}$ &  \\
                2b-b& 19.1 & 2.88 &  6.6 $\times10^{-4}$ &  \\
                2b-c& 191.0 & 2.88 &  6.6 $\times10^{-3}$ &  \\
                \hline\hline
                \end{tabular}
\caption{Grid of DALI chemical models. Each model starts with atomic abundances and is run in time-dependent mode with a stopping time of 10$^6$ yr. }\label{tab:grid}
\end{table}

\begin{figure*}[!t]
        \centering
        \includegraphics[width=18cm]{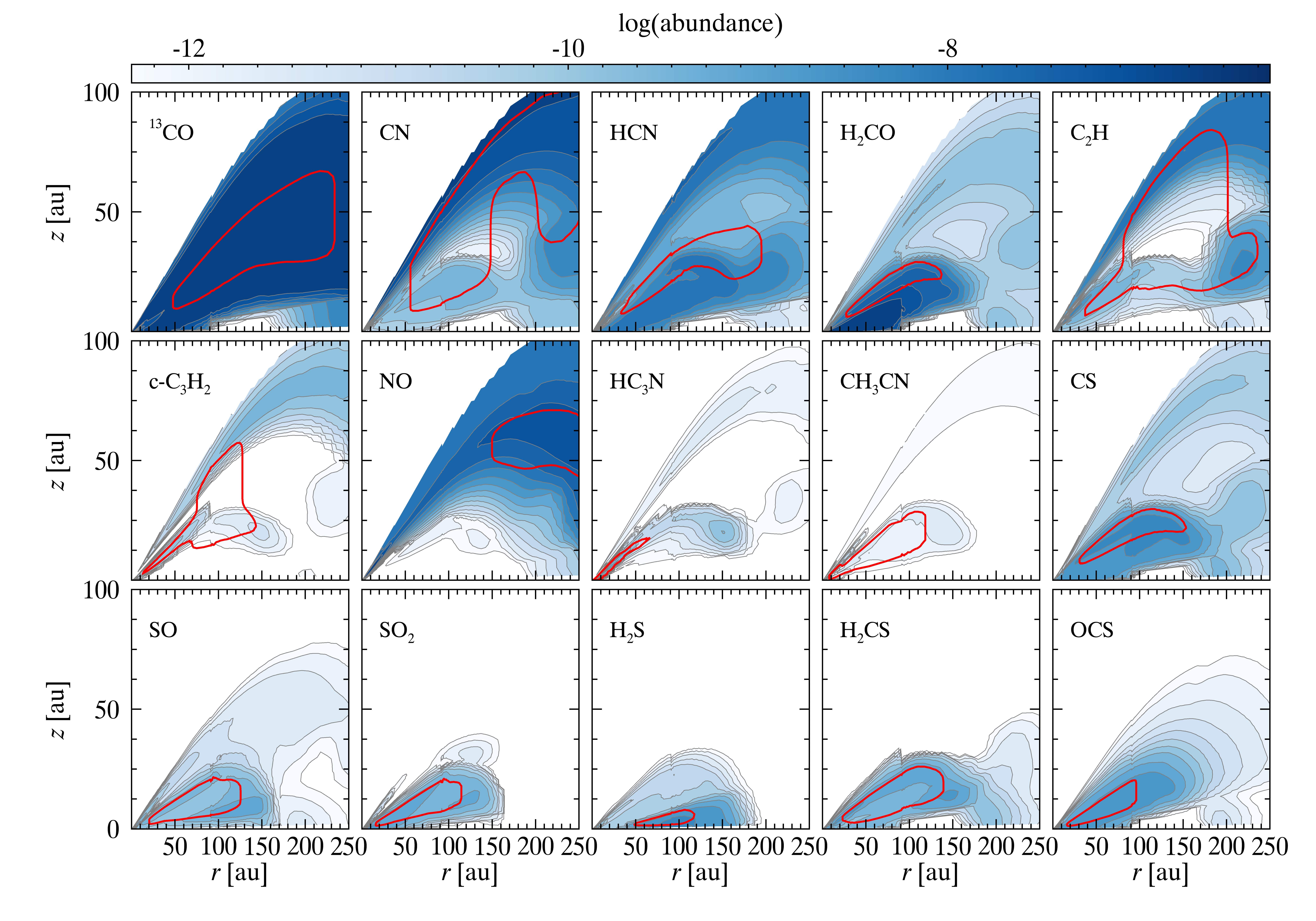}
        \caption{Abundance structure of the molecules studied in this paper for the reference model (boldface values in Table~\ref{tab:setup} and initial elemental abundances as in model 2b in Table~\ref{tab:grid}). The (red) solid line indicates the layer including 75\% of the emission for the transitions listed in Table~\ref{tab:lines}.}\label{fig:dali2}
\end{figure*}

\begin{table}[!t]
\centering
        \begin{tabular}{llll}
                \hline\hline
                Species & Transition & E$_u$ (K) & $\nu$ (GHz) \\
                \hline
                $^{13}$CO      & $2-1$                                       & 15.87 & 220.39868 \\
                CN                  & $3-2$                                      & 32.66 & 340.24777 \\
                HCN                & 4-3                                           & 42.53 & 354.50547 \\
                H$_2$CO        & $3_{0,3} - 2_{0,2}$                  & 62.50 & 218.21000\\
                C$_2$H           & $J=\frac{7}{2}-\frac{5}{2}, F=4-3$                             & 25.15 & 262.00426\\
                c-C$_3$H$_2$ & $6_{1,6} - 5_{0,5}$                & 38.61 & 217.82215\\
                HC$_3$N         & $27-26$                                 & 165.04 & 245.60632 \\ 
                CH$_3$CN      & $12_2 - 11_2$                       & 97.44 & 220.73027 \\
                CS                   & 5-4                                          & 35.30 & 244.93556 \\
                SO                  & $7_8 - 6_7$                             & 81.20 & 340.71415 \\
                NO                  &    $J=\frac{7}{2}-\frac{5}{2}, F = \frac{9}{2}-\frac{7}{2}$                                             & 36.13 & 351.04352 \\
                OCS                & 18-17                                      & 99,81 & 218.90335\\
                SO$_2$           & $5_{3,3} - 4_{2,2}$                & 35.90 & 351.25722\\
                H$_2$S          & $2_{2,0} - 2_{1,1}$                  &  84.00 & 216.71044 \\ 
                H$_2$CS       & $7_{16} - 6_{15}$                    & 60.0 & 244.04850\\
                \hline\hline
        \end{tabular}
        \caption{List of molecular transitions analyzed here. The collisional rates are taken from the LAMDA database (\citealt{Schoier05}; https://home.strw.leidenuniv.nl/~moldata/). Note in particular that the C$_2$H rates have been recently updated with those from \citet{Dagdigian18} for collisional partners ortho and para H$_2$. }\label{tab:lines}
\end{table}

\subsection{Model grid}
A grid of models was created for different initial abundances of N, O, and S relative to H while keeping the carbon abundance 
fixed to C/H=1.35 $\times 10^{-4}$. 
Table~\ref{tab:grid} reports the list of models with the initial abundances and the 
corresponding elemental abundance ratios C/O, C/N, N/O, and S/O. The initial abundances of the other elements 
are fixed and equal to: He/H = 7.59 $\times$ 10$^{-2}$, Mg/H = 4.17 $\times$ 10$^{-7}$, Si/H = 7.94 $\times$ 10$^{-6}$, Fe/H = 4.27 $\times$ 10$^{-7}$. 
Four different values of oxygen abundances have been explored in order to investigate a wide range of C/O, from 0.046 to 1.562 . 
Multiple sets of models have been created with the same ranges of C/O but for three abundances of nitrogen with the C/N spanning a range of 0.63 to 63.8. 
A subset of models for different sulfur abundances was computed.

\begin{figure*}[!t]
        \centering
        \includegraphics[width=18cm]{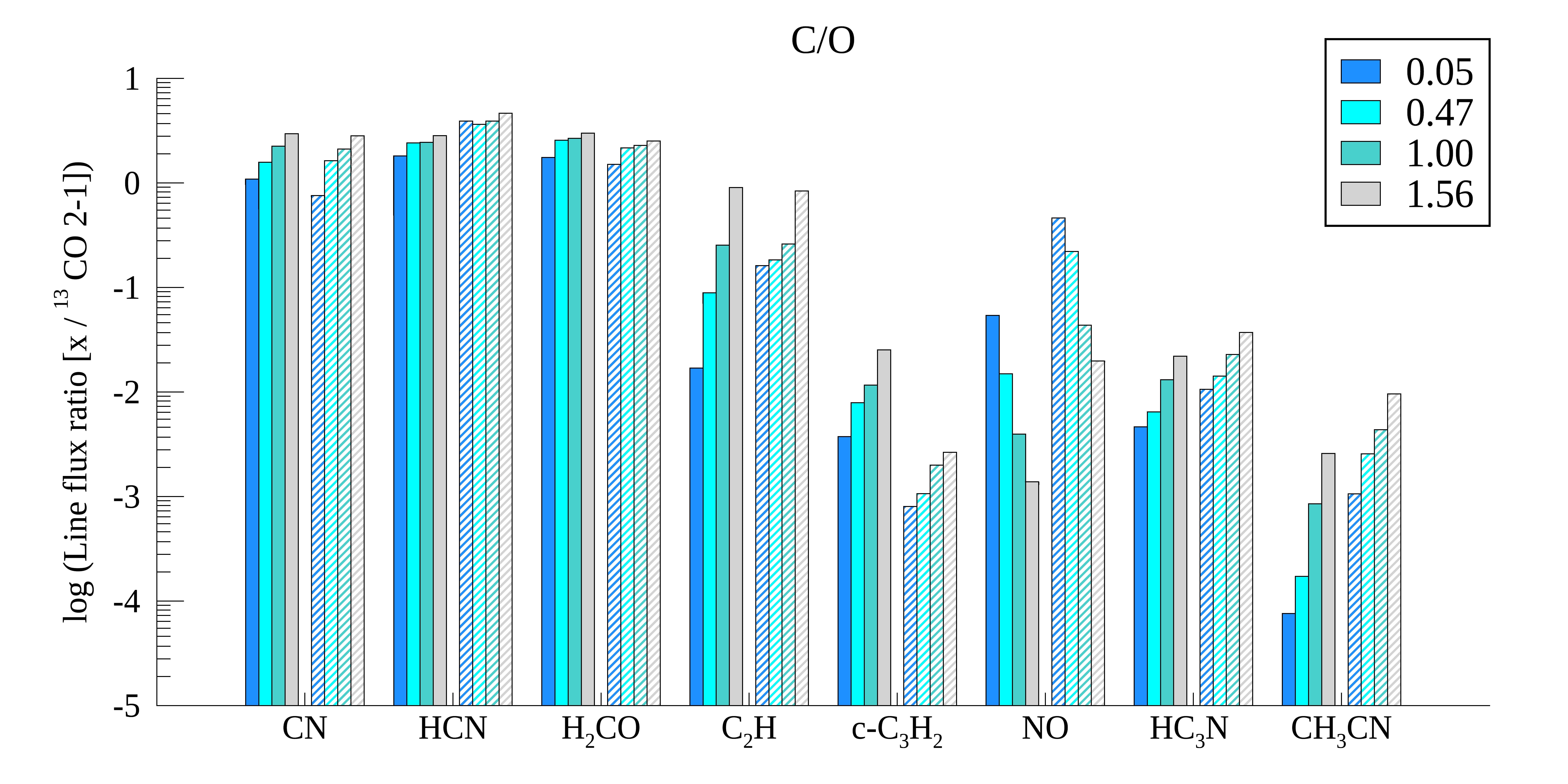}
        \caption{Line flux ratios as a function of the initial  C/O ratio for the transitions listed in Table~\ref{tab:lines}. 
        In all cases, the abundances of carbon, nitrogen, and sulfur are fixed at $1.35 \times 10^{-4}$, $2.14 \times 10^{-5}$ and $1.91 \times 10^{-8}$, respectively 
        (model ID 2a, 2b, 2c, and 2d from Table~\ref{tab:grid}).}\label{fig:CO}
\end{figure*}

\smallskip
\noindent
The input physical parameters of the reference model are listed in Table~\ref{tab:setup} (values in boldface). Additional models have 
been created to analyze the response of the molecular transitions to the input physical structure for a given set of elemental abundances. 
Among the others, the parameters that influence the line intensities are the disk-flaring and scale height, the gas-to-dust mass ratio ,
and the dust properties \citep[minimum grain size and dust settling, e.g.,][]{Fedele16}. Indeed, $\psi$ and $h_{\rm c}$ control the gas temperature (hence line excitation and intensity),
while the dust properties affect the opacity. The incident X-ray luminosity and total stellar luminosity are also investigated here.  
For each of these parameters, three different values are being examined here (Table~\ref{tab:setup}).

\smallskip
\noindent
In this work, we selected a set of carbon-, nitrogen- and sulphur-bearing species previously detected
in disks at millimeter wavelengths with, for example, ALMA and NOEMA: CN, HCN, NO, C$_{2}$H, c--C$_{3}$H$_{2}$, H$_{2}$CO, HC$_{3}$N, CH$_{3}$CN, CS, SO, H$_{2}$S, and H$_{2}$CS \citep[e.g.,][]
{Dutrey97,Qi08a,Qi13a,Qi13b,Henning10,Chapillon12,Guilloteau13,Oberg15,Loomis18,Bergin16,Bergner18,Phuong18,Podio19,Legal19}.
To our knowledge, NO has not been detected in protoplanetary disks yet, while it was detected in protostellar envelopes and shocks \citep[e.g.,][]{Codella18}.

\smallskip
\noindent
The density and temperature structures for the reference model are shown in Figure~\ref{fig:dali1}. Figure~\ref{fig:dali2} shows the abundance structures of 
the molecules studied here (initial elemental abundance as in model 2b of Table~\ref{tab:grid}) along with the emitting layer of the transitions of Table~\ref{tab:lines}.

\begin{figure*}[!t]
        \centering
        \includegraphics[width=18cm]{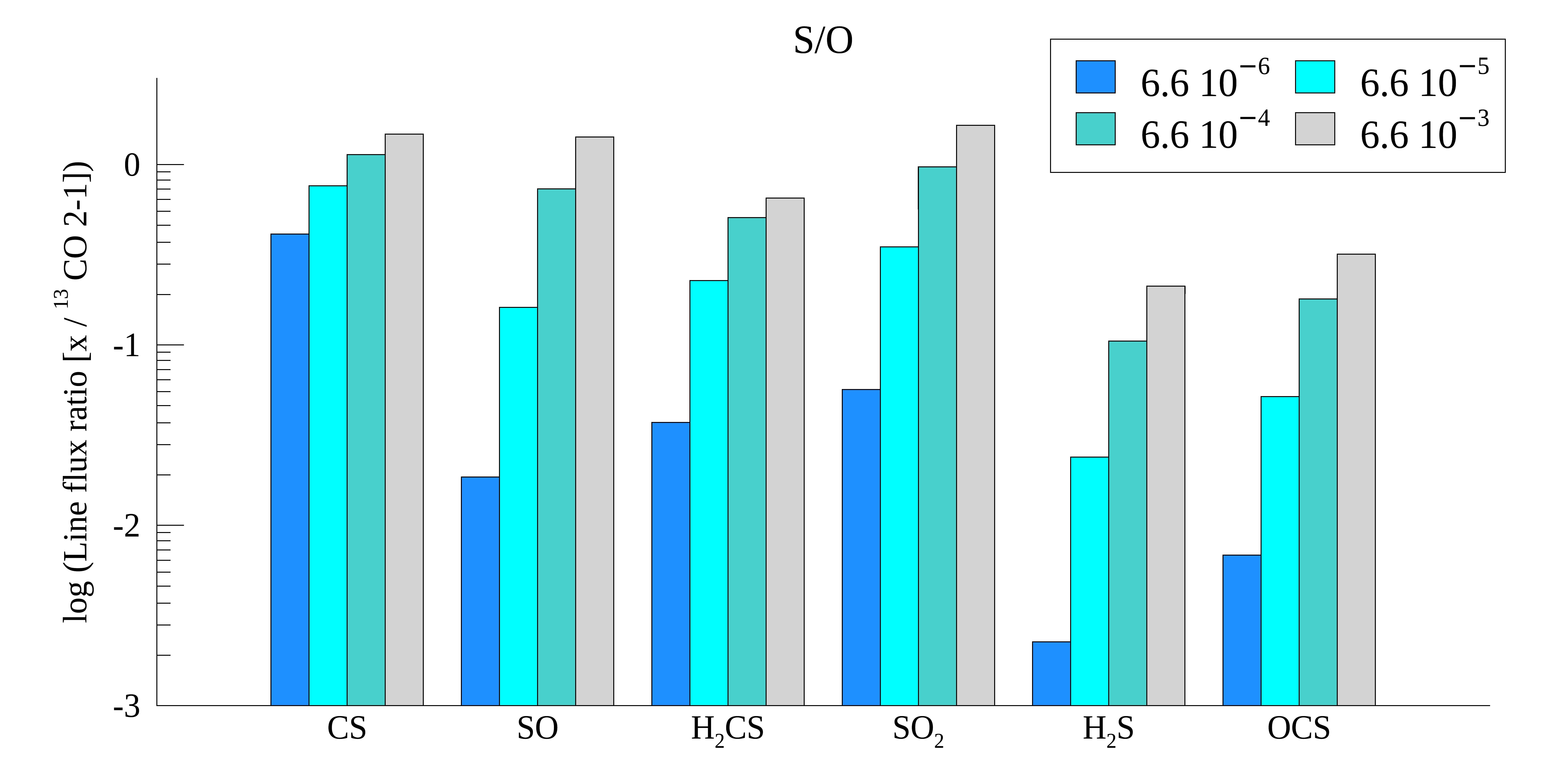}
        \caption{Line flux ratios of sulphur-bearing species as a function of the initial S/O abundance ratio (Table~\ref{tab:grid}, ID 2b-a, 2b, 2b-c, 2b-d). In this case, only the sulphur-bearing species are shown.}\label{fig:SO}
\end{figure*}

\section{Results}\label{sec:results}
Disk-integrated line fluxes are computed for a selection of commonly detected molecular transitions (Table~\ref{tab:lines}). 
The lines are selected with the aim of minimizing the frequency settings with ALMA. In particular,
all these transitions can be observed with three frequency settings (two in band six and one in band seven).
In all cases, the disk inclination is fixed at 30$^{\circ}$ and the distance is set to 100\,pc. 
The flux of the individual transitions is sensitive to the excitation conditions and molecular abundances. 
To first order, the $^{13}$CO $J=2-1$ transition is a proxy of the temperature gradient of the disk. Thus, with the aim
of normalizing the effect of the temperature structure on the line excitation, we divide all the line fluxes by the flux of the 
$^{13}$CO $J=2-1$ line. As a consequence, the changes in the line flux ratios reflect the intrinsic variation of the molecular
abundances. 

\subsection{Line flux ratios versus elemental abundance ratios}
The line flux ratios are shown in Figure~\ref{fig:CO} as a function of the initial C/O abundance ratio and for two different values of the 
initial abundance of nitrogen: filled bars are the models with low nitrogen abundance (model IDs: 1a, 1b, 1c, and 1d in Table~\ref{tab:grid}), 
while the dashed bars show the high N abundance case (model IDs 3a, 3b, 3c, and 3d in Table~\ref{tab:grid}). Further models are presented in the Appendix.
All the species show a positive trend with the initial C/O abundance ratio. The only exception is NO, whose line-flux ratio decreases with increasing C/O.  
Among the others, NO, C$_2$H, c-C$_3$H$_2$, HC$_3$N, and CH$_3$CN are the most sensitive species to the gas-phase C/O abundance ratio, with the flux ratio increasing by 1-2 orders of magnitude. 

\smallskip
\noindent
The models with the high nitrogen abundance show the same trends with regard to the C/O value. Increasing the nitrogen abundance has the 
effect of increasing the flux ratios of HCN, NO, HC$_3$N, and CH$_3$CN. Interestingly, the ratio CN/$^{13}$CO does not change substantially. 
Notably, the c-C$_3$H$_2$/$^{13}$CO ratio decreases drastically compared to the low-nitrogen case. 

 \smallskip
 \noindent
Figure~\ref{fig:SO} shows the behavior of sulphur-bearing species as a function of the initial S/O abundance ratio. In this case, the nitrogen, 
carbon, and oxygen abundances are fixed (model IDs 2b, 4b, 5b). We note that initial sulphur abundance is varied by two orders of magnitude to account
for the low detection rate of sulphur species in disks \citep[e.g.,][]{Semenov18}. The flux ratio of SO and H$_2$S varies by almost the same amount.

\subsection{Line fluxes versus stellar-disk physical parameters}
\begin{figure*}[!t]
        \centering
        \includegraphics[width=9cm]{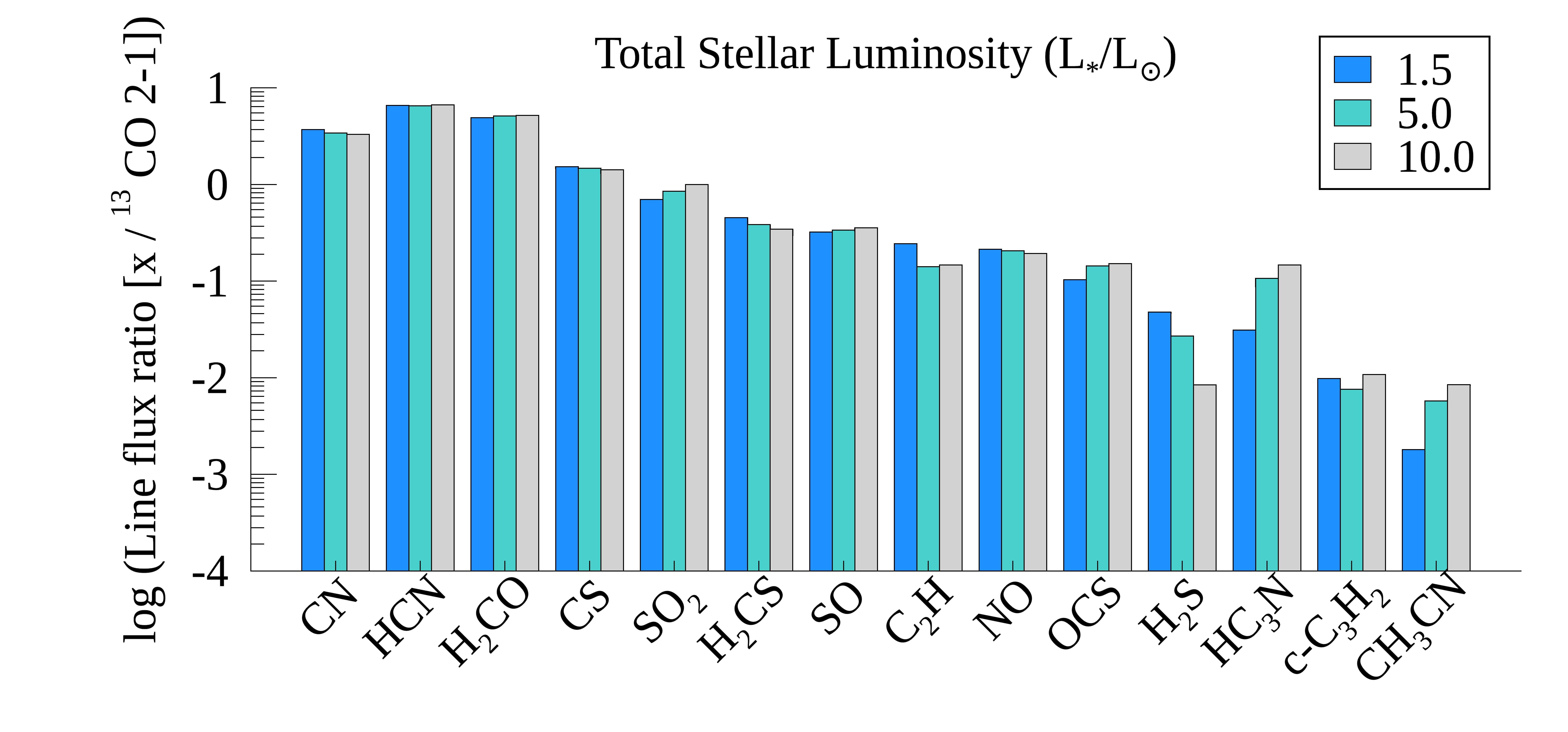}
        \includegraphics[width=9cm]{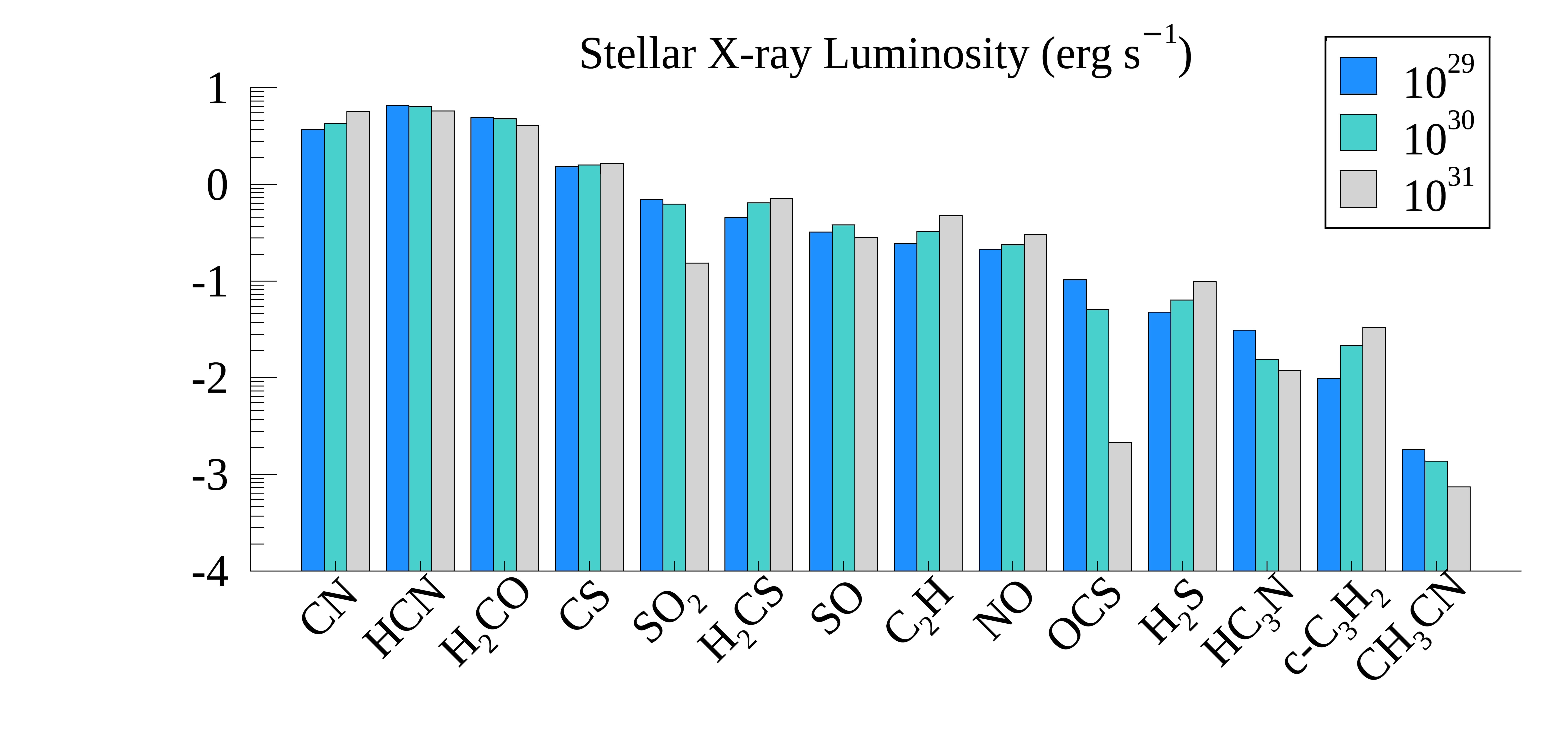}
        \includegraphics[width=9cm]{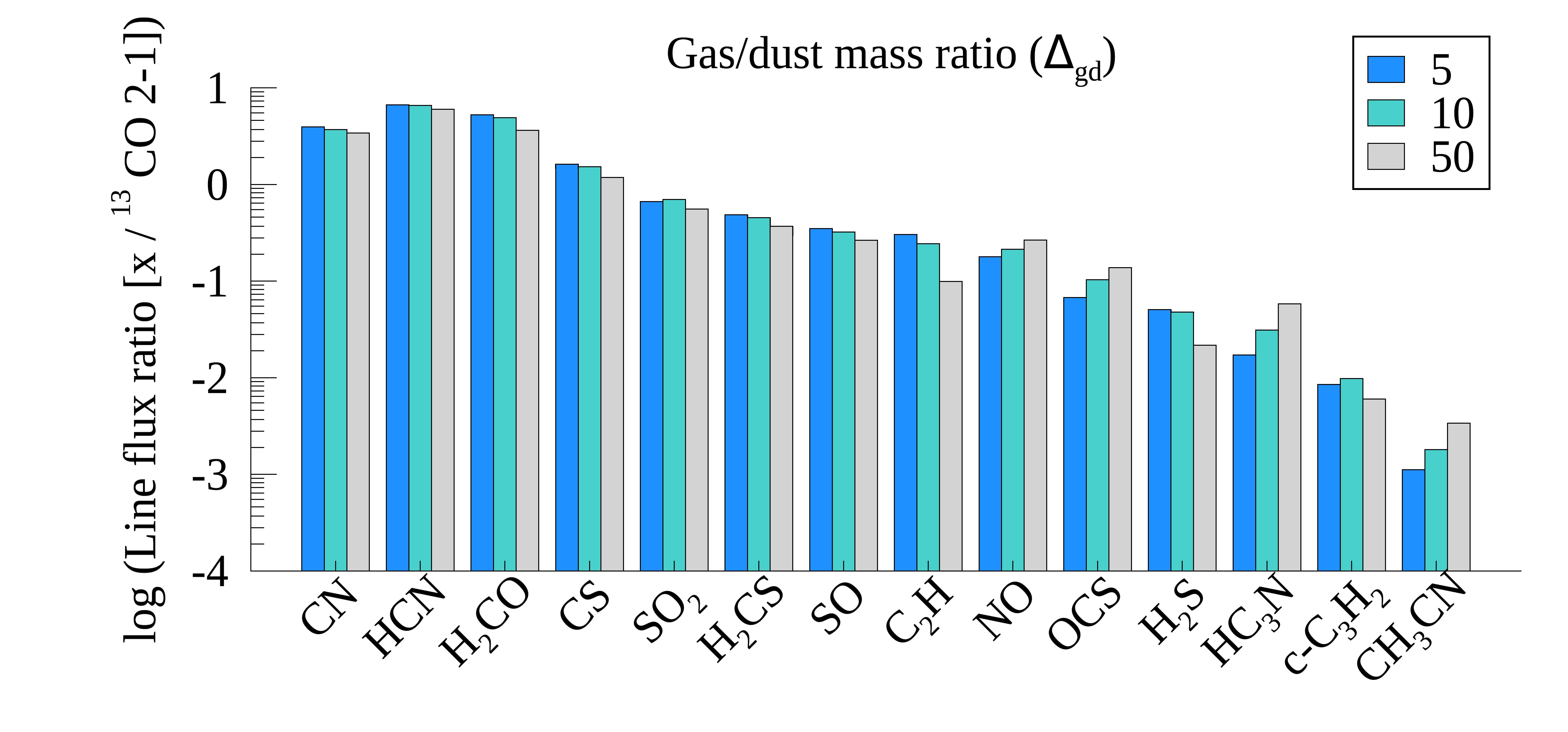}
        \includegraphics[width=9cm]{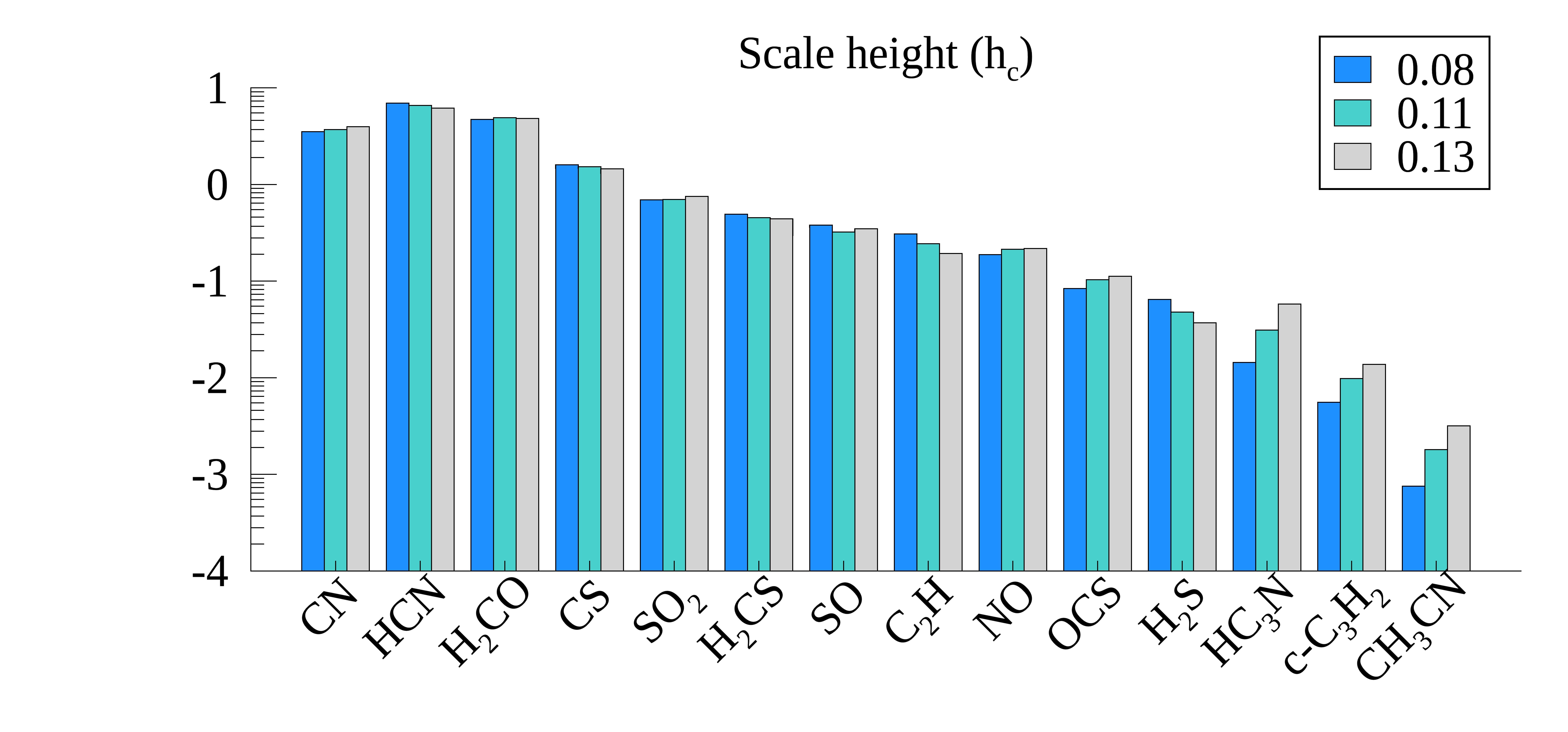}
        \includegraphics[width=9cm]{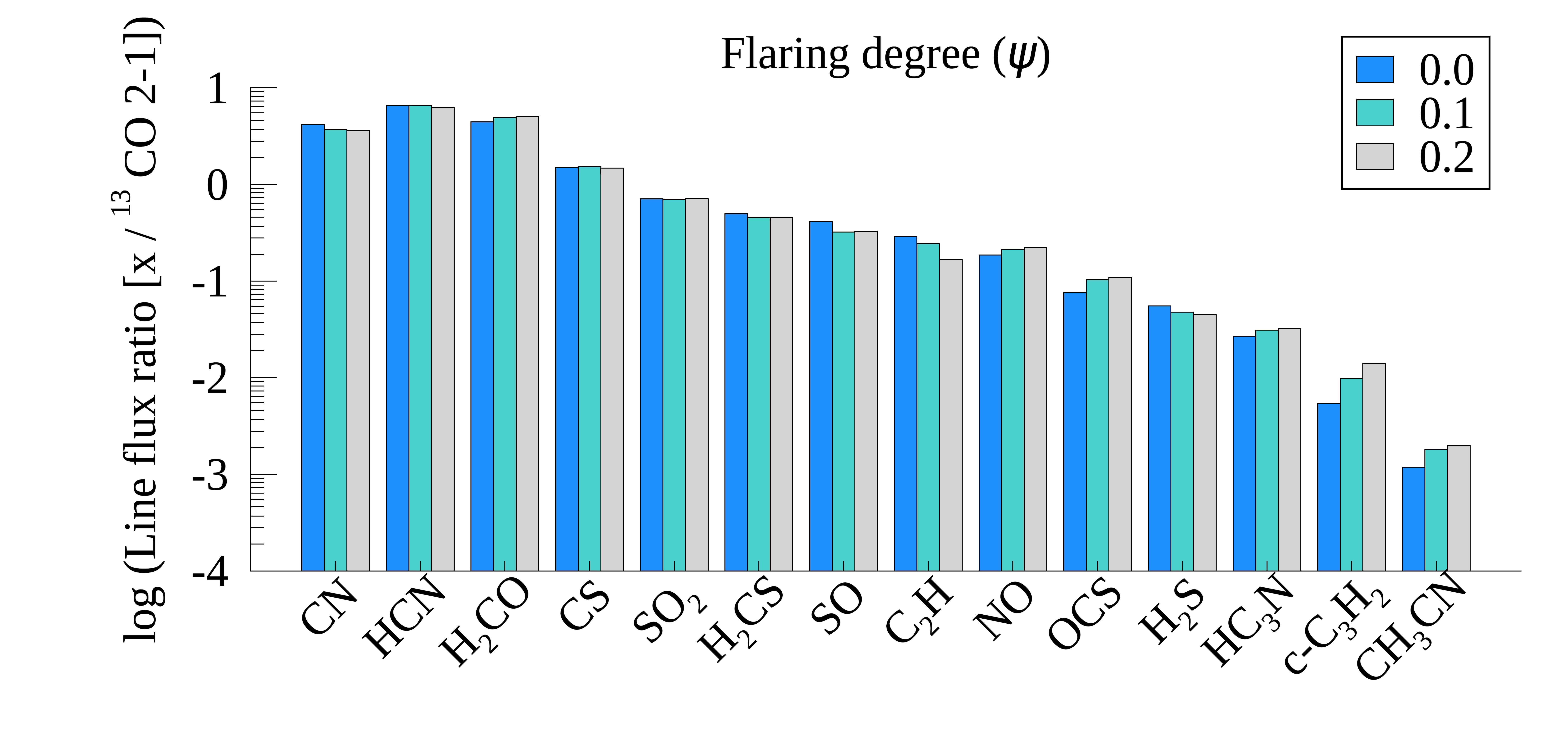}
        \includegraphics[width=9cm]{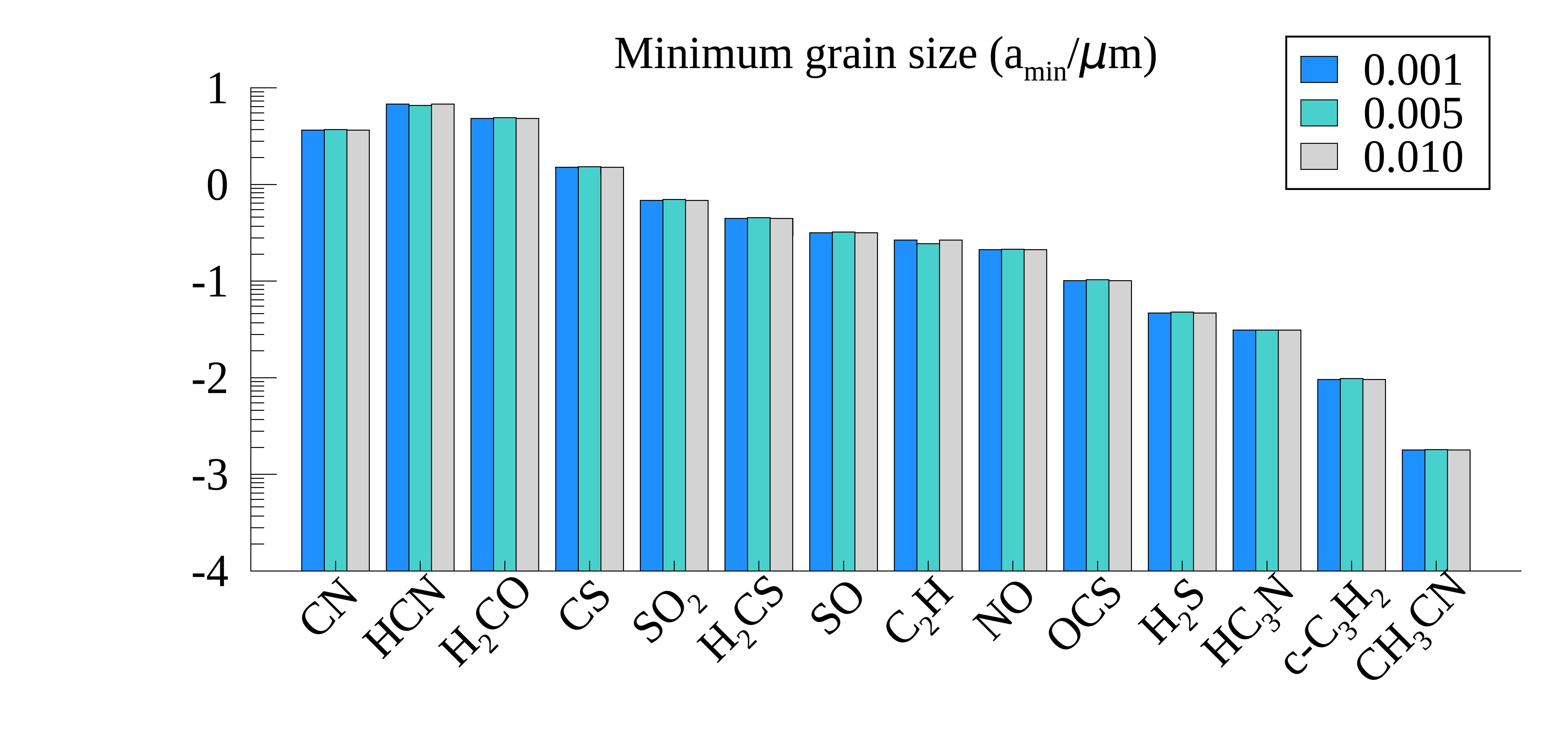}
        \includegraphics[width=9cm]{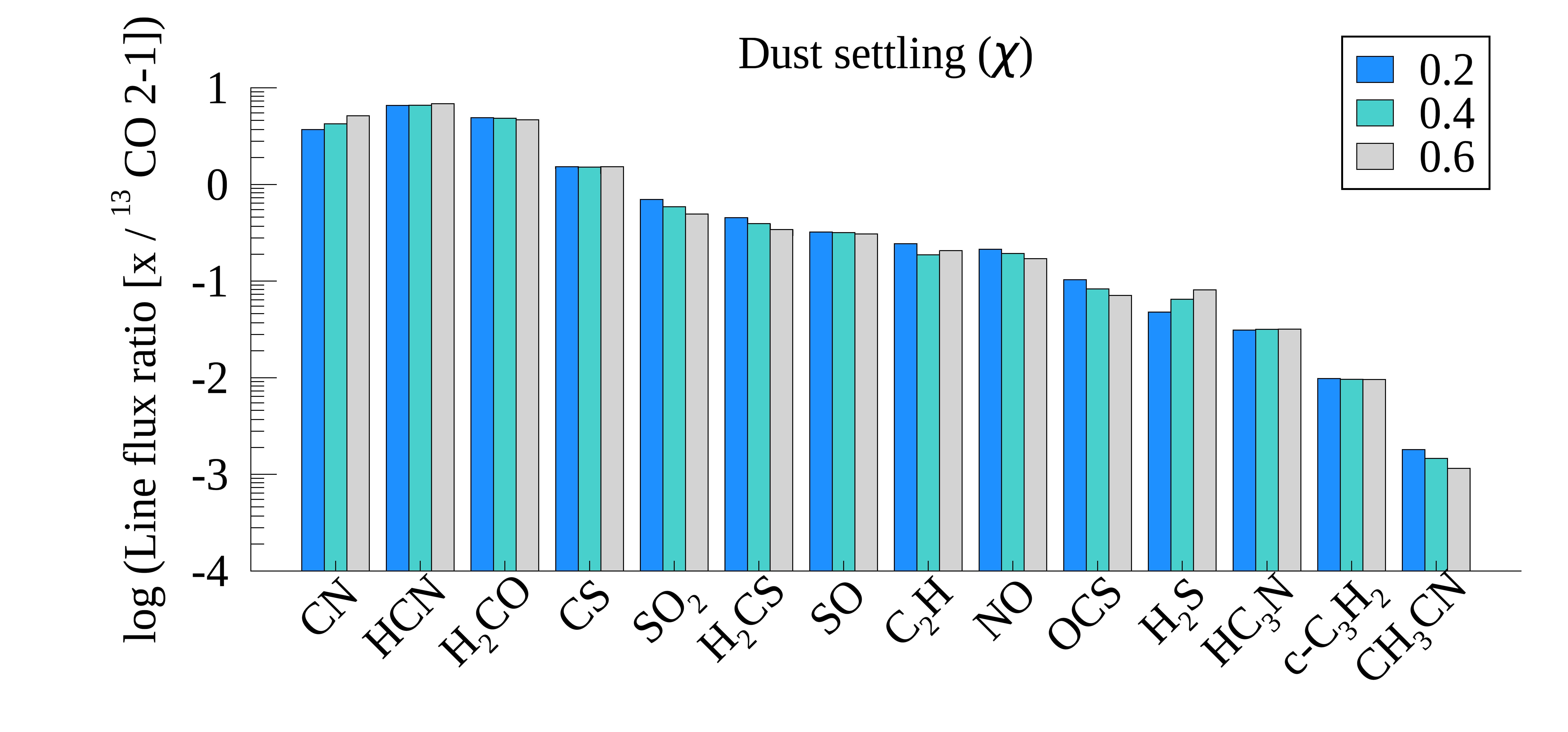}
        \includegraphics[width=9cm]{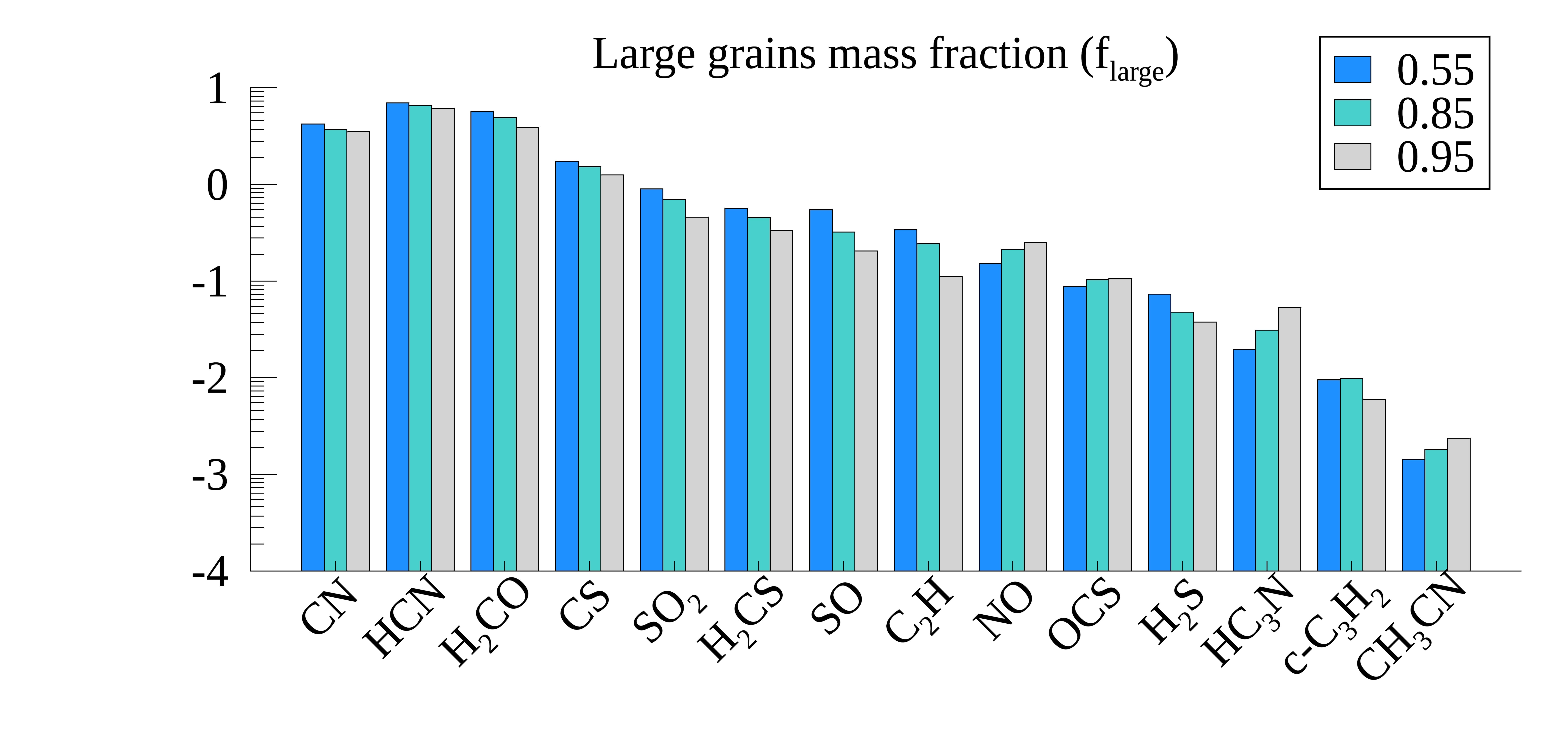}
        \caption{Line flux ratios for different stellar and disk physical properties (see Table~\ref{tab:setup} for definitions). In all cases, the initial elemental abundances are fixed as in model run 2b in Table~\ref{tab:grid}. }\label{fig:grid}
\end{figure*}

The line-flux ratios presented above can also be sensitive to the stellar and disk physical parameters that alter the temperature and 
density structures, hence the excitation conditions. The results of the physics model grid are shown in Figure~\ref{fig:grid}. In all cases, the initial elemental
abundances are fixed in the same way as in the reference model (model ID 2b in Table~\ref{tab:grid}). 
A first immediate result is that the flux ratios of some species remain almost unaltered in all cases: this is the case of CN, HCN, H$_2$CO, CS, and H$_2$CS. 
On the other hand, variations by a factor of a few are predicted for the other species. The effect of the individual parameters are
described as follows:
({\it i}) Stellar luminosity ($L_*$) - the flux ratio of HC$_3$N and CH$_3$CN increases by a factor of a few when the luminosity goes from 1.5 to 10\,L$_{\odot}$;
({\it ii}) Stellar X-ray luminosity ($L_X$) - varying $L_X$ by two orders of magnitude induces only minor changes in the line-flux ratios, SO$_2$ and OCS decrease substantially for large $L_X$ values, on the contrary, H$_2$S and c-C$_3$H$_2$ slightly increase with increasing $L_X$;  
({\it iii}) Gas-to-dust mass ratio ($\Delta_{gd}$) - variations by a factor of a few are predicted for C$_2$H (decreasing with increasing $\Delta_{gd}$), HC$_3$N, and CH$_3$CN (increasing with $\Delta_{gd}$);
({\it iv}) Disk scale height ($h_c$) - increasing $h_c$ lead to higher flux ratios for C$_3$H$_2$, HC$_3$N, and CH$_3$CN;
({\it v}) Disk flaring ($\psi$) - minor changes are predicted for c-C$_3$H$_2$, whose line-flux ratio increases slightly with $\psi$, no substantial changes are expected for the other transitions;
({\it vi}) Minimum grain size ($a_{min}$) - varying $a_{min}$ from 0.001\,\micron ~ to 0.01\,\micron, it has no impact on the line-flux ratios;
({\it vii}) Dust settling ($\chi$) - minor changes are predicted for some sulphur-bearing species, as well as for CH$_3$CN;
({\it viii}) Large grain mass fraction ($f_{large}$) - increasing the mass fraction of the large grain population from 55\% to 95\% has the effect of lowering the line-flux ratio of several species by a factor of a few. On the contrary, NO, HC$_3$N, and CH$_3$CN show an opposite trend.

\section{Discussion}\label{sec:discussion}
For a detailed description of the molecular composition of protoplanetary disks, we invite the reader to read the seminal papers of, for example,
\citet{Aikawa02}, \citet{Semenov11}, \citet{Henning13}, \citet{Walsh15}, \citet{Guilloteau16}, \citet{Agundez18}, \citet{Legal19}. This section aims instead to discuss the chemical routes 
that lead to the chemical enrichment as a function of the elemental abundance ratios. The following discussion is based on reactions taken from the UMIST chemical network \citep{Woodall07}  
(see Section \ref{sec:model}).

\smallskip
\noindent
An interesting finding of our modeling (see Section~\ref{sec:results}) is that the initial C/O value simultaneously influences several species, that emit within different layers of the disk.
The higher abundance of the hydrocarbons C$_2$H and c-C$_3$H$_2$ with C/O is a direct consequence of the enhanced carbon chemistry \citep{Bergin16}.
Indeed, inside a protoplanetary disk there are multiple paths for the formation of C$_2$H starting from neutral atomic and molecular carbon. 
The formation routes starting with atomic carbon are described in \citet{Bergin16} and \citet{Bergner19} and are not repeated here. 
A third important route is:

\begin{align}
&\mathrm{C_2 + H_3^+ \rightarrow C_2H^+ + H_2,}     \\
&\mathrm{C_2H^+ + H_2 \rightarrow C_2H_2^+ + H,}   \\
&\mathrm{C_2H_2^+ + e^- \rightarrow C_2H + H.}
\end{align}

\smallskip
\noindent
We also note that C$_2$H$^+$ can further react to form C$_2$H$_2^+$:

\begin{align}
&\mathrm{C_2H^+ + CH_4 \rightarrow C_2H_2^+ + CH_3,}  \\
&\mathrm{C_2H^+ + HCN \rightarrow C_2H_2^+ + CN.}\label{eq:2}
\end{align} 
 
\smallskip
\noindent
Similarly, the gas-phase formation of c-C$_3$H$_2$ in disks proceeds from the C$_3$H$^{+}$: 

\begin{align}
&\mathrm{C_2H_2^+ + H_2 \rightarrow C_2H_3^+ + H}  ,\\
&\mathrm{C_2H_2^+ + C^+ \rightarrow C_3H^+ + H} \label{eq:9}    ,\\
&\mathrm{C_2H_3^+ + C \rightarrow C_3H^+ + H_2} \label{eq:10} ,\\
&\mathrm{C_3H^+ + H_2 \rightarrow C_3H_3^+ } ,\\
&\mathrm{C_3H_3^+ + e^- \rightarrow C_3H_2 + H.}
\end{align}

\smallskip
\noindent
Although  reaction~\ref{eq:10} is about an order of magnitude slower than reaction~\ref{eq:9}, it might still influence the production of C$_3$H$^{+}$.\\
In turn, the higher content of hydrocarbons trigger the production of the nitriles. The formation of HC$_3$N is indeed strictly linked to 
C$_2$H as the main reactions are:

\begin{align}
&\mathrm{C_2H + HCN \rightarrow HC_3N + H,}  \\
&\mathrm{C_2H + HNC \rightarrow HC_3N + H,}  \\
&\mathrm{CN + C_2H_2 \rightarrow HC_3N + H,}  \\
&\mathrm{C_3H_2 + N \rightarrow HC_3N + H,}\label{eq:3}  \\
&\mathrm{C_2H + CN \rightarrow HC_3N + h\nu,} 
\end{align}

\smallskip
\noindent
while that of the methyl cyanide CH$_3$CN proceeds primarily from: 

\begin{align}
&\mathrm{CH_3^+ + HCN \rightarrow  CH_3CNH^+},\\
&\mathrm{CH_3CNH^+ + e^- \rightarrow CH_3CN +  H,}
\end{align}

\smallskip
\noindent
and potentially (but less probably) from:

\begin{align}
&\mathrm{CH_3 + CN \rightarrow CH_3CN.}
\end{align}

\smallskip
\noindent 
Thus, the formation of both HC$_3$N and CH$_3$CN  (strictly linked to the presence of hydrocarbons) requires free atomic or 
molecular carbon to be triggered. This likely explains the positive trend of the line-flux ratios of the nitriles transitions with C/O. \\
However, the situation is the opposite for NO, which is destroyed by atomic carbon and hydrocarbons:

\begin{align}
&\mathrm{C + NO \rightarrow CN + O} \nonumber  \\
&\mathrm{~~~~~~~~~~~~~~\rightarrow CO + N,}  \\
&\mathrm{CH + NO \rightarrow HCN + O} \nonumber \\
&\mathrm{~~~~~~~~~~~~~~~~~ \rightarrow CN + OH} \nonumber \\
&\mathrm{~~~~~~~~~~~~~~~~~ \rightarrow HCO + N,}  \\
&\mathrm{CH_2 +NO \rightarrow HCN + OH.} 
\end{align}

\smallskip
\noindent
The trend of the nitriles with the the initial nitrogen abundance can be easily understood.
An interesting result is the behavior of C$_2$H and c-C$_3$H$_2$ with N/H (and N/O):
in the first case, the line-flux ratio increases slightly with the nitrogen abundance, and this is due to the formation of C$_2$H$_2^+$ 
(precursor of C$_2$H) via reaction~\ref{eq:2}. On the other hand, c-C$_3$H$_2$ is destroyed (among others) by the reaction 
with atomic nitrogen (reaction~\ref{eq:3}) to form HC$_3$N (depending on the location in the disk). 

\smallskip
\noindent
Finally, the behavior of the sulphur-bearing species with S/O is easily understood, as their abundance increases with increasing 
abundance of sulphur. Among the species studied, SO and H$_2$S are more sensitive to the global sulphur abundance compared 
to CS and H$_2$CS. The low detection rate of SO and H$_2$S \citep[e.g.,][]{Booth18, Semenov18} and the detection of CS and H$_2$CS 
\citep[e.g.,][]{Legal19a} hints at a low gas--phase abundance of sulphur in disks, with sulphur being primarily locked in 
refractory elements \citep[e.g.,][]{Kama19}. In this regard, it is interesting to note that the direct 
comparison of H$_2$CO and H$_2$CS can provide direct information on the S/O abundance ratio in disks as the main formation 
route of both species share the same parent molecule, the radical CH$_{3}$:  

\begin{align}
&\mathrm{CH_3 + O \rightarrow H_2CO + H,} \\
&\mathrm{CH_3 + S \rightarrow H_2CS + H.} 
\end{align} 

\smallskip
\noindent
From the discussion above, it is clearly evident that the simultaneous analysis of multiple molecular transitions is a powerful tool to 
constrain the elemental abundance ratios in protoplanetary disks. A proper selection of molecular lines  is further important to breaking the 
degeneracy between the elemental abundances and physical parameters. For example, the line-flux ratios of c-C$_3$H$_2$, HC$_3$N 
and CH$_3$CN increases with increasing C/O ratio, but also with the disk scale height ($h_c$); therefore, one could not distinguish 
between the two. This is not the case for C$_2$H, CN, HCN, and H$_2$CO, for example, which do not substantially vary with $h_c$. We also note 
that the situation is similar for the flaring degree.

\subsection{Considerations about dynamical processes}
The results presented in this paper are based on a static disk model. 
Viscous accretion, inward migration of pebbles, and vertical mixing may induce a variation of the elemental abundance ratios 
\citep[e.g.,][]{Piso15,Oberg16, 
Booth17, Krijt18}. As a consequence, the molecular abundances and the line fluxes also vary accordingly. 
The significance of radial and vertical mixing depends on several factors such as the viscosity, the ionization rate (for viscous accretion),
 the degree of turbulence (for the vertical mixing), and the coupling of gas and dust (radial drift). We note in particular, that large ($\gtrsim 1\,$mm) dust grains 
can be easily trapped in local pressure maxima \citep[e.g.,][]{Pinilla12, Zhu14}, slowing down the inward migration or ice-coated pebbles.

\noindent
Nevertheless, the molecular transitions investigated in this paper are mostly sensitive to the cold gas reservoir in the outer disk. 
The molecular content of the inner disk region (spatial scales $\lesssim 10\,$au) can be traced via infrared spectroscopy, which is sensitive to
warm and hot gas.  
In particular, future observations with JWST will also allow us to detect several 
species highly sensitive to the C/O ratio (e.g., C$_2$H$_2$, CH$_3$, CH$_4$, C$_3$H$_4$, C$_6$H$_6$), enabling us to constrain 
the molecular richness and the elemental abundance ratio. A direct comparison between infrared and millimeter molecular
transitions can thus provide information about the radial distribution of the elemental abundance ratios.

\section{Conclusions}\label{sec:conclusions}
The results presented in this paper demonstrate that the gas-phase elemental abundance ratio of C/O, N/O, and S/O in protoplanetary 
disks can be constrained by means of line-flux ratios of multiple molecular transitions. The stellar and disk physical properties appear to only slightly change the flux ratios studied here, and only some species are affected. This further demonstrates that the flux of the $^{13}$CO $J=2-1$ line is 
indeed a good proxy of the physical conditions in disks. Nevertheless, our study strongly suggests the need to simultaneously compare the flux ratios of multiple species to 
discern between physical properties and elemental abundances.

\noindent
The advent of broad-band correlators in (sub)millimeter interferometry offers a unique opportunity to carry out simultaneous observations
of multiple species. This allows us to perform a statistical investigation of the chemical composition of  disks and to determine
the elemental abundance ratios. Such a study is important to make the link between the atmospheric composition of planets and the primordial 
composition of protoplanetary disks.

\begin{acknowledgements}
DF acknowledges financial support from the Italian Ministry of Education, Universities and Research, project SIR (RBSI14ZRHR). 
CF acknowledges {\it i)} financial support from the French National Research Agency in the framework of the Investissements d'Avenir program (ANR-15-IDEX-02), through the funding of the "Origin of Life" project of the Univ. Grenoble-Alpes, and {\it ii)}, funding from the European Research Council (ERC) under the European Unions Horizon 2020 research and innovation programme, for the Project {\it The Dawn of Organic Chemistry} (DOC), grant agreement No 741002.
We thank the referee (Alex Cridland) for the fruitful comments and suggestions.
\end{acknowledgements}

\bibliographystyle{aa} 
\bibliography{mybib} 
%

\begin{appendix}
\section{Model with carbon depletion}

\begin{figure*}[th!]
        \centering
        \includegraphics[width=18cm]{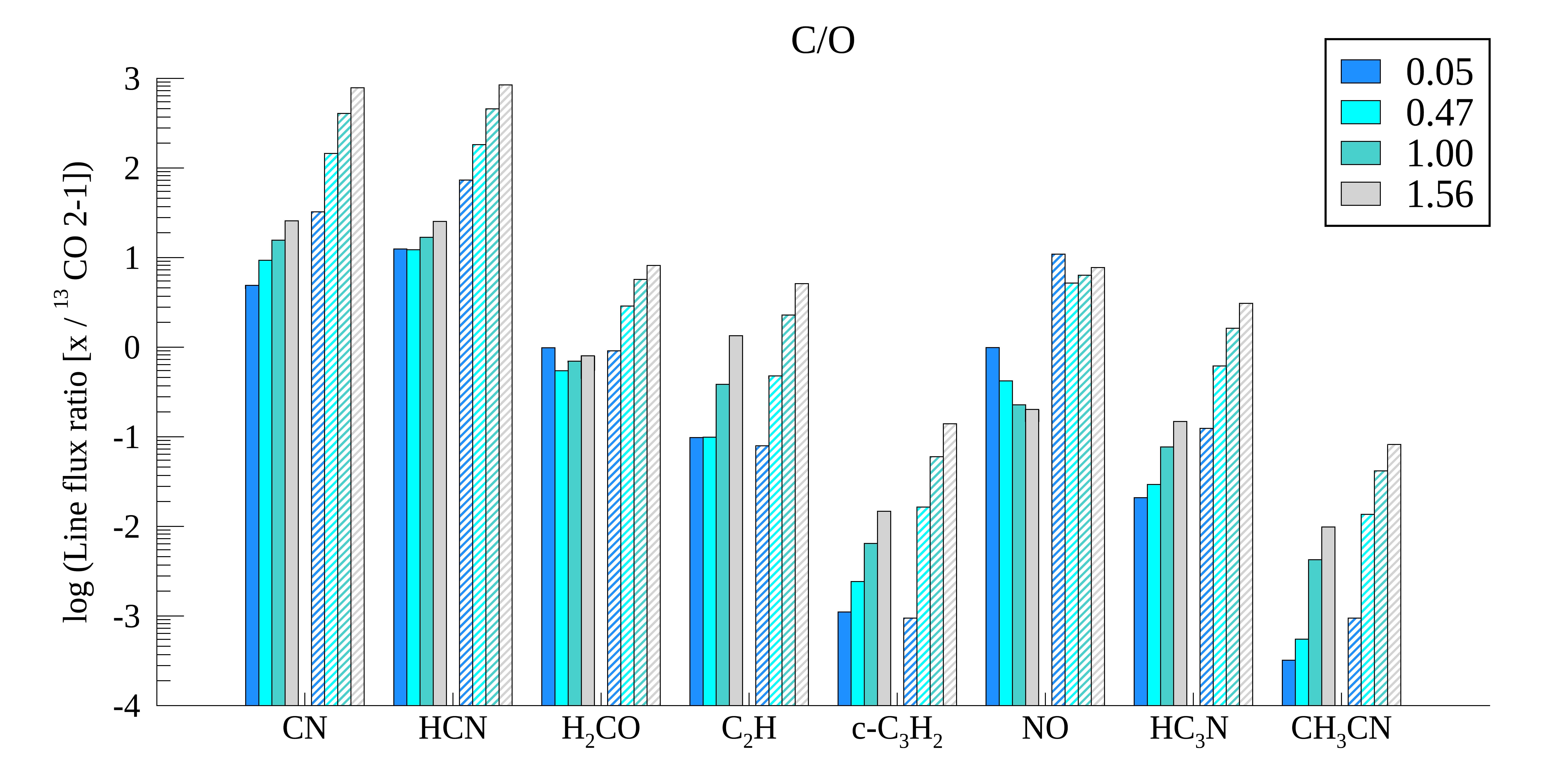}
        \caption{Same as Figure~\ref{fig:CO} for C/H = 1.35 $\times 10^{-5}$ (filled bars) and 1.35 $\times 10^{-6}$ (dashed bars) (see Table~\ref{tab:appendix})}\label{fig:CO_2}
\end{figure*}

A direct link may exist between the depletion of different elements in disks. This depends on their volatility with nitrogen, carbon, oxygen, and sulphur, forming a sequential path from the most to the least volatile element that we can observe. While there is evidence of depletion of C, N, and O in disks, as of today there is no evidence of depletion of nitrogen \citep[e.g.,][]{Cleeves18}.
We thus performed further DALI models fixing the nitrogen abundance and lowering the elemental abundance of carbon (Table~\ref{tab:appendix}). The results are shown in Figure~\ref{fig:CO_2}.
Overall, the line-flux ratios show similar trends as in Figure~\ref{fig:CO}. Notably, the transitions of the nitrogen-bearing species are now much stronger than $^{13}$CO $J=2-1$.

\begin{table}[!h]
\centering
        \begin{tabular}{lllll}
                \hline\hline
                ID & N/H & O/H & C/O & N/O \\
                \#     & $\times 10^{-5}$ & $\times 10^{-4}$ & &  \\
                \hline
                        & \multicolumn{4}{c}{C=1.35 $\times$ 10$^{-5}$}\\
                \hline
                        & N/H & O/H & C/O & N/O \\
                \#     & $\times 10^{-5}$ & $\times 10^{-4}$ & &  \\
                \hline
                4a& 2.14 & 28.8 & 0.047 & 7.4 $\times10^{-4}$  \\
                4b& 2.14 & 2.88 & 0.469 & 7.4 $\times10^{-3}$  \\
                4c& 2.14 & 1.35 & 1.000 & 1.6 $\times10^{-2}$  \\
                4d& 2.14 & 0.86 & 1.562 & 2.5 $\times10^{-2}$  \\       
                \hline\hline
                        & \multicolumn{4}{c}{C=1.35 $\times$ 10$^{-6}$}\\
                \hline
                        & N/H & O/H & C/O & N/O \\
                \#     & $\times 10^{-5}$ & $\times 10^{-4}$ & &  \\
                \hline
                5a& 2.14 & 28.8 & 0.047 & 7.4 $\times10^{-4}$  \\
                5b& 2.14 & 2.88 & 0.469 & 7.4 $\times10^{-3}$  \\
                5c& 2.14 & 1.35 & 1.000 & 1.6 $\times10^{-2}$  \\
                5d& 2.14 & 0.86 & 1.562 & 2.5 $\times10^{-2}$  \\       
                \hline\hline
                \end{tabular}
\caption{Further DALI chemical models for two different values of the carbon abundance.}\label{tab:appendix}
\end{table}

\end{appendix}

\end{document}